\documentclass[aps,amsmath,amssymb,prapplied,reprint]{revtex4-2}
\usepackage{CJK}
\usepackage{dcolumn}
\usepackage{bm}
\usepackage{siunitx}
\usepackage{float}
\usepackage{color}
\usepackage[normalem]{ulem}
\usepackage{pgfplots}
\usepackage{graphicx}
\pgfplotsset{compat=1.16}
\graphicspath{{figs/}} 

\usepackage{hyperref}
\hypersetup{
	colorlinks=true,
	linkcolor=blue,
	anchorcolor=blue,
	citecolor=blue,
	urlcolor=blue,
}

\newcommand {\dif}{\mathop{}\!\mathrm{d}}
\newcommand {\weim}{\,\mu\text{m}}
\newcommand {\weis}{\,\mu\text{s}}
\newcommand {\etal}{\textit{et al.}}
\newcommand{\avg}[1]{\left\langle #1 \right\rangle}

\begin{document}

	\begin{CJK*}{UTF8}{gbsn}
		\preprint{11\#202B}
		\title{Implementation of two-dimensional selective acoustic tweezers merely using four straight interdigitated transducers:a numerical proof of concept of radiation field synthesis by pulsed acoustic waves}
		
		\author{Shuhan Chen 
		}
		\author{Jia Zhou 
		}
		\email{jia.zhou@fudan.edu.cn}
		
		\author{Antoine Riaud}		
		\email{antoine.riaud@gmail.com}
              \altaffiliation[current affiliation: ]{ABB Corporate Research Center, Baden-D{\"a}ttwil, Switzerland}	
		\affiliation{School of Microelectronics, Fudan University, Shanghai 200433, China}

		\date{\today}
		
		\begin{abstract}
			Selective acoustic tweezers can focus the acoustic radiation force on a single particle to manipulate it without affecting its neighbors. This has long required highly complex hardware. In this numerical study, we show that pulsed acoustic waves can be used for the selective manipulation of particles using only two pairs of orthogonal transducers. While these tweezers are well-known for their ability to manipulate arrays of particles, we show that selectivity can be achieved by using sequences of acoustic pulses to iteratively construct a combined acoustic potential focused only on the target particle.  
		\end{abstract}
		
		\maketitle
	\end{CJK*}
\section{\label{intro}Introduction}
Acoustic tweezers manipulate particles and cells without contact using the acoustic radiation force (ARF). At equal power density, they provide a trapping force 100,000 times larger than optical tweezers\cite{thomas2017acoustical}. This confers them a high biocompatibility which makes them promising for a range of biological applications\cite{cui2023robust}. In some of these applications, large amounts of cells need to be manipulated or clumped together, which is usually achieved using two pairs of orthogonal transducers to create a tunable standing acoustic field\cite{guo2014controlling,yang2022harmonic}. In other applications, it is important to capture only one particle among many other identical ones. Such selective manipulation requires to finely craft the acoustic radiation force field to act only on a single particle, which is one of the longstanding challenges of acoustofluidics.

Traditional selective manipulation methods rely on acoustic vortices (a class of acoustic fields featuring a helical wavefront) that have the advantage of having a vanishing pressure on their propagation axes while having a maximum intensity around this axis\cite{baudoin2019folding,baudoin2020spatially}. These fields can be synthesized using transducer arrays or holographic structures such as spiraling transducers. Transducer arrays are difficult and costly to miniaturize, while holograms must be mechanically translated to move the trap.

\begin{figure}[!htbp]
    \includegraphics[width=\columnwidth]{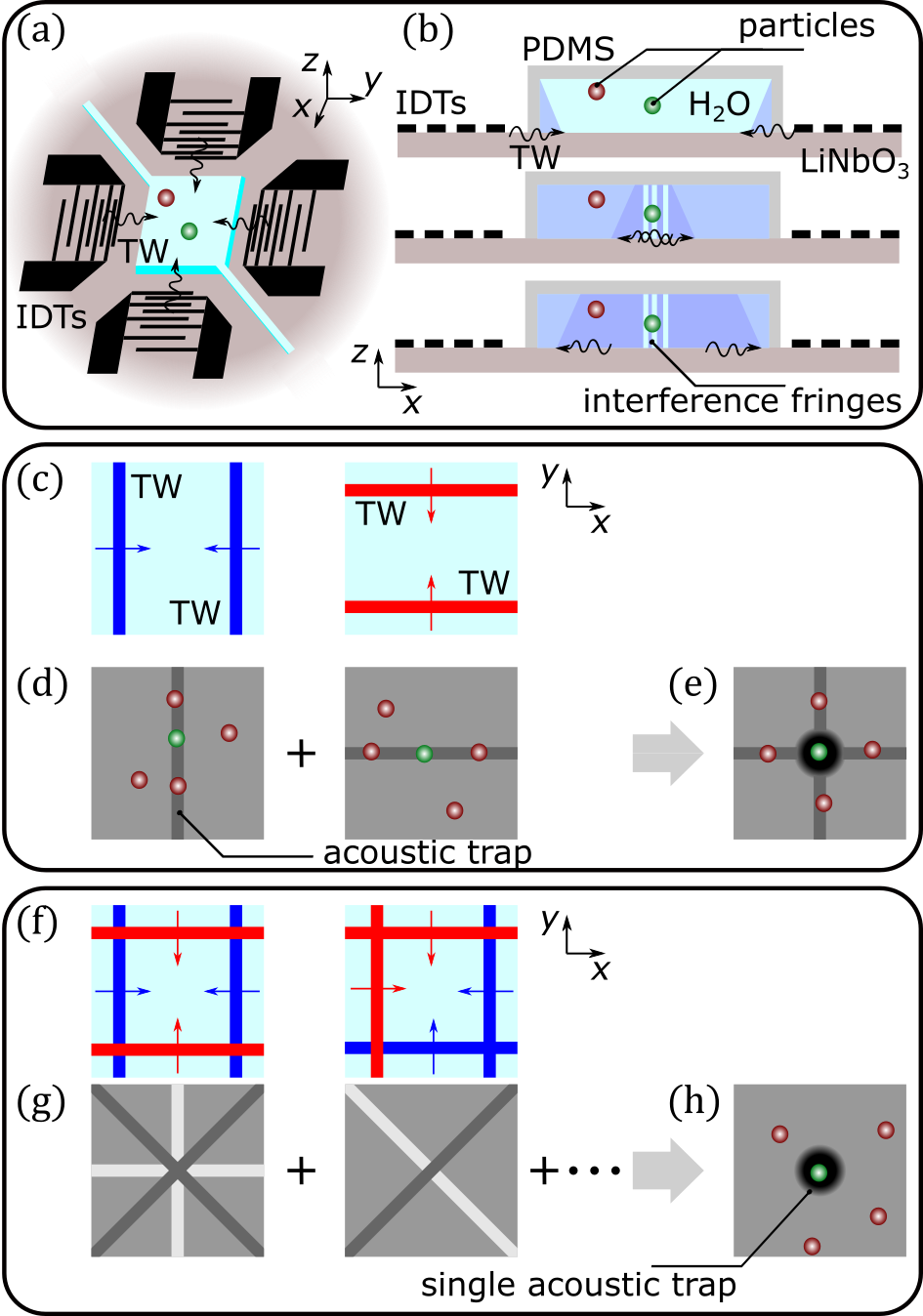}
    \caption{\label{fig:overview}Principle for selective two-dimensional trapping using time multiplexing. \textbf{(a) tweezers overview:} The acoustic tweezers combine two pairs of IDTs facing each other. The IDTs send pulsed traveling waves (TW) to a microfluidic chamber located at the center. The chamber contains the particles that should be manipulated. \textbf{(b) manipulation principle:} The traveling waves propagate guided by the substrate (LiNbO$_3$) and create a leaky Rayleigh wave in the fluid (darker blue). When the two waves meet at the center of the channel, destructive interference creates a localized pattern of fringes that can capture particles. The manipulation is selective because only the green particle (in the fringed region) is trapped. \textbf{(c) from one-dimensional to two-dimensional manipulation by time-multiplexing:} while using one pair of IDTs creates a one-dimensional acoustic trap that allows trapping particles along a line \textbf{(d)}, energizing one pair of transducers after another creates two one-dimensional acoustic traps that can be superimposed \textbf{(e)} to create a combined two-dimensional cross-shaped acoustic trap. However, this trap also captures unwanted particles (in red).\textbf{(f) selective manipulation using all four transducers:} energizing all four transducers with combinations of positive or negative excitation yields a larger set of acoustic traps \textbf{(g)}, which can be superimposed to create a combined acoustic potential with a single trap \textbf{(h)}.}
\end{figure}

Pulsed acoustic waves are a promising alternative to helical wave fields. Generation of pulses using miniaturized interdigitated transducers were developed when interdigitated transducers (IDTs) were used for pulse compression in radar technologies using chirp IDTs \cite{royer1999elastic} or slanted finger IDTs combined with dispersive grating reflectors for compression ratios exceeding 20 (\cite{williamson1973use}). Beyond acoustofluidics, these technologies might have found a new life in quantum processing \cite{wang2022generation}. Both technologies have been used in acoustofluidics \cite{song2022design}, with chirp IDTs used to control inter-particle distance in acoustic manipulation \cite{guo2014controlling}, and slanted-finger IDTs for particle sorting \cite{li2021physical} and fine-positioning of acoustic fields relative to an acoustofluidic device \cite{schmid2014sorting} or droplet \cite{khalid2013control}. 

In one-dimension, Collins \etal \cite{collins2016acoustic} have used a single pair of IDTs to trap particles at a predefined location of a microchannel. Using such chirp IDTs to increase the transducer bandwidth, Xu \etal \cite{xu2021fourier} has devised a frequency-multiplexing method able to achieve $\lambda/4$ resolution for pulsed manipulation. The same year, Wang \etal\ \cite{wang2021acoustic} has provided a rigorous theoretical foundation for the manipulation of small particles using pulsed acoustic waves, which can be numerically implemented on finite element software \cite{chen2023numerical}. Despite this rapid progress, pulsed acoustic tweezers have been limited to one-dimensional manipulation (excluding expensive laser-guided acoustic tweezers \cite{wang2022laser}). This is because acoustic fields along each dimension create a one-dimensional acoustic trap, which results in a cross-pattern (see for instance Fig. 4 of Xu \etal \cite{xu2021fourier}) instead of a single point trap. Therefore, any particle along the cross will be captured and selectivity is lost.

In this paper, we show that using sequences of pulses allows combining multiple acoustic radiation potential fields, some being plus-shaped $+$ and others being cross-shaped $\times$ (Fig.~\ref{fig:overview}.g). This time-division multiplexing (TDM) enables the synthesis of a combined acoustic potential field having a single potential well surrounded by a smooth potential landscape (Fig.~\ref{fig:overview}.h) which eventually allows selective manipulation.

The paper is structured as follows: after summarizing the theory of ARF by pulsed acoustic waves and presenting the numerical model, we show the synthesis of a single potential well at various locations, followed by the simulation of particle manipulation. We then simulate the selective manipulation of particles along a square path.

\section{\label{theory}Theory}

The particle dynamics in time-multiplexed acoustic tweezers cover three different timescales: fast nearly-periodic oscillations over an acoustic period ($<0.1\weis$), gradual build-up of momentum during repeated irradiation by a set of acoustic sub-fields ($\approx80\weis$), and slow motion of the particle itself towards the acoustic trap created by the combination of sub-fields ($\approx \SI{1}{\s}$).

In the following, we consider particles of density $\rho_p$ and compressibility $\kappa_p$ and radius $R_p$ immersed in an inviscid fluid of density $\rho_0$, velocity of sound $c_0$, compressibility $\kappa_0=1/(\rho_0 c_0^2)$. In the simulations, we assume the microchannel height to be small compared to the acoustic wavelength. Indeed, when the channel height exceeds half a wavelength, the acoustic field exhibits pressure nodes in the vertical direction, which degrades selectivity because multiple traps exist in the $z$-dimension \cite{bernard2017controlled}. These particles are irradiated by a pulsed acoustic field of wavelength $\lambda_0>>R_p$, having an incident pressure field $\tilde{p}_{\text{in}}$ and incident vibration velocity field $\tilde{\mathbf{v}}_{\text{in}}$. The resulting ARF reads \cite{wang2021acoustic}:
\begin{equation}  
\mathbf{F}_{\mathrm{rad}}=-\frac{4\pi R_{p}^3}{3} \nabla  \avg{\mathcal{U}}, \label{eq: Frad}
\end{equation}
with $\mathcal{U}$ a mathematical artifact that can be interpreted as the instantaneous Gor'kov potential:
\begin{equation}
    \mathcal{U} = \frac{f_1}{2\rho_0 c_0^2} \tilde{p}_{\text{in}}^2 - \frac{3f_2 \rho_0}{4} {\tilde{\mathbf{v}}_{\text{in}}}^2, \label{eq: U} 
\end{equation}
where $\avg{\mathcal{U}}$ indicates the time-average of the instantaneous Gor'kov potential $\mathcal{U}$, and $f_1=1-\kappa_0 / \kappa_p$ and $f_2=2(\rho_p-\rho_0)/(2\rho_p+\rho_0)$ are the monopole and dipole scattering coefficients, respectively.

Computing $\mathcal{U}$ requires knowing the incident acoustic pressure field $\tilde{p}_{\text{in}}$ and incident vibration velocity field $\tilde{\mathbf{v}}_{\text{in}}$. 

Resolving the acoustic field as a superposition of plane monochromatic waves $\tilde{p} = \int_{-\infty}^\infty \int_{\mathbb{R}^3} \hat{p}(\omega)\text{e}^{\text{i}(k_x x + k_y y + k_z z)-\text{i}\omega t} \dif \mathbf{k} \text{d} \omega$ (using the Fourier transform in space and time), we note that for each of these plane waves satisfies $\rho_0 \text{i}\omega\hat{v}_z = -\text{i} k_z \hat{p}$, that is $\hat{v}_z =\hat{p} / \rho_0\omega k_z $, with $k_z$ given by the dispersion relation $k_z = \sqrt{k_{0}^2 - k_{x}^2 - k_{y}^2}$. Continuity boundary conditions at the liquid-solid interface ensure  $k_x=k_{x,\text{SAW}}$ and $k_y=k_{y,\text{SAW}}$. For the surface acoustic wave (SAW), the dispersion relation reads $k_{x,\text{SAW}}^2 + k_{y,\text{SAW}}^2=\omega^2 / c_\text{SAW}^2$. Substituting in the equation of $k_z$, we get $k_z = \sqrt{k_0^2 -\left( \omega / c_{\text{SAW}}\right)^2}$, which can be recast as $k_z = k_0 \cos{\Theta_R}$, with the Rayleigh angle $\Theta_R=\arccos(c_0 / c_{\text{SAW}})$. Substituting in the expression for the velocity, we get $\hat{v}_z = \hat{p}/ (\rho_0 c_0 \cos{\Theta_R}) $, which is now independent of the frequency. However, it is still dependent on the propagation direction because in general, SAW propagation velocity $c_{\text{SAW}}$ depends on the propagation direction. Yet, this anisotropy is only of a few percent \cite{riaud2015taming} and will be neglected here. Under this assumption, the relation $\hat{v}_z = \hat{p}/ (\rho_0 c_0 \cos{\Theta_R}) $ becomes independent of the direction, and allows using the well-known acoustic impedance to deduce the $z$-component of the velocity from the pressure field in the transient direct-space domain: 
\begin{equation}
\tilde{v}_{\text{in},z} = \frac{1}{\rho_0 c_0 \cos{\Theta_R}} \tilde{p}_{\text{in}}. \label{p_z}
\end{equation}

We note that the factor $1/\cos{\Theta_R}$ was mentioned in previous works on radiation force by monochromatic waves\cite{simon2017particle}. The shallow channel assumption allows computing the incident pressure field using the continuity condition  $\tilde{v}_{\text{in},z} = \tilde{v}_{\text{SAW},z}$. Neglecting attenuation, the SAW obeys the transient two-dimensional d'Alembert equation in the $x$-$y$ plane, in which we substitute Eq.~\eqref{p_z}:
\begin{equation}
\frac{1}{c_{\text{SAW}}^2} \frac{\partial^{2}\tilde{p}_{\text{in}}}{\partial t^{2}} = \nabla^{2} \tilde{p}_{\text{in}}, \label{eq: DAlembert}
\end{equation}
where the factor $\rho_0 c_0 \cos{\Theta_R}$ was simplified on both sides. The $x$ and $y$ components of $\tilde{\mathbf{v}}_{\text{in}}$ can be deduced from the acceleration field $\tilde{\mathbf{a}}_{\text{in}}=-(1/\rho_0)\nabla \tilde{p}_{\text{in}}$ by integration over time $\tilde{\mathbf{v}}_{\text{in}} = \int \tilde{\mathbf{a}}_{\text{in}} \dif t$. In summary:
\begin{subequations} 
\begin{align} 
\tilde{\mathbf{v}}_{\text{in}}=-\int \frac{1}{\rho_0}\nabla \tilde{p}_{\text{in}} \dif t, \label{eq: vxvy} \\
\text{along $x$ and $y$, and:} \notag \\
    \tilde{v}_{\text{in},z} = \frac{1}{\rho_0 c_0 \cos{\Theta_R}} \tilde{p}_{\text{in}}, \label{eq: vz} 
\end{align} 
\label{eq:v}
\end{subequations}

As illustrated in Fig.~\ref{fig:overview}(g), selective manipulation is achieved by cycling through a set of $n$ sub-fields to generate a combined potential $\mathcal{U}_\text{comb} = \avg{\mathcal{U}} = (1/T) \int_{0}^{T} \mathcal{U} \dif t $, where $T$ is the multiplexing period. Assuming that these fields are sufficiently well separated in time such that they do not overlap in the microchannel (and that the particle does not ring between sub-fields\cite{wang2021acoustic}), we get $ $
\begin{equation}
    \mathcal{U}_\text{comb} =\frac{1}{T}\sum_{j=1}^{n} \int_{0}^{T/n} \mathcal{U}_j \dif t= \frac{1}{n}\sum_{j=1}^{n} \avg{\mathcal{U}_j} \label{eq:u_eff}.
\end{equation}


The particle trajectory is computed by assuming that the ARF is balanced by the Stokes drag force and the Coulomb dry friction force (between the particle and the microchannel surface):
\begin{equation}
    \mathbf{0} = \mathbf{F}_{\mathrm{rad}} - \textbf{F}_d - \textbf{F}_c,  \label{eq:v_p} 
\end{equation}
where $\textbf{F}_d = 6\pi \mu R_p\textbf{v}_p$ and $\textbf{F}_c=\min(F_{\mathrm{rad}},F_{c,\mathrm{max}})\textbf{f}$, with $F_{\mathrm{rad}}=||\mathbf{F}_{\mathrm{rad}}||_2$ the magnitude of the ARF and $\mathbf{f}=\mathbf{F}_{\mathrm{rad}}/F_{\mathrm{rad}}$ the direction of the ARF. Consequently, when the ARF is small, the Coulomb dry friction force balances it exactly. As the force exceeds the maximum Coulomb force ($F_{c,\mathrm{max}}$), the excess is balanced by the drag force. In this instance, the fluid viscosity is set to $\mu=\SI{1}{mPa\cdot s}$. The Coulomb force is set to $F_{c,\mathrm{max}}=0.22$\SI{}{\nano\newton}, and the particle diameter $d_p=10\weim$.

\section{\label{method}Method}

In our simulations, the ARF of each sub-field is first computed using a two-dimensional acoustic model, then the combined force field is computed and exported to a Python script to compute the particle motion. For dynamic manipulation, the combined field is updated at preset regular intervals during the particle motion by loading successive combined fields from a dataset. Each combined field is made of 4 sub-fields, and a typical manipulation such as the one shown at the end of the paper requires approximately 48 combined fields.

\subsection{Acoustic model}
The transient ARF of the model will be computed via the numerical approach proposed by Chen \etal \cite{chen2023numerical} with the commercial finite-element-method software COMSOL Multiphysics version 5.4.  

\underline{Implementation:} The computation procedure for any given sub-field follows Chen \etal\cite{chen2023numerical} method:
\begin{enumerate}
    \item the acoustic pressure field is computed according to Eq. \eqref{eq: DAlembert}, 
    \item from the pressure field, we deduce the velocity field (Eq. \eqref{eq:v}), 
    \item knowledge of acoustic pressure and velocity allows computing the instantaneous potential (Eq. \eqref{eq: U}).
\end{enumerate}

This procedure is repeated for each sub-field constituting a combined field. Then, the contribution of the sub-fields are summed according to Eq.~\eqref{eq:u_eff}. Finally, the particle motion is computed in Python by solving Eq. \eqref{eq:v_p}.
 
The ARF is exported as the orthogonal $x$-component $F_x = -(4\pi R_{p}^3/3) \partial_x \mathcal{U}_\text{comb}$ and $y$-component $F_y = -(4\pi R_{p}^3/3) \partial_y \mathcal{U}_\text{comb}$ unless otherwise specified. 

\underline{Parameters:} The simulation domain is a square of side-length $W_d=\SI{1}{\mm}$ and centered on the origin of the coordinates. The velocity of sound $c_{\text{SAW}}=\SI{4000}{\m\per s}$, the density $\rho_{\text{SAW}}=\SI{4640}{\kg\per\m^{3}}$; for particles, the radius $R_p=5\weim$; The density of fluid medium solutions is often adjusted to be close to the density of small spheres, in order to balance buoyancy and gravity. We note that while close to balance, it is not possible in practice to balance gravity and buoyancy exactly, and the particles are actually at rest against either the floor or the ceiling of the microfluidic chamber. Balancing density results in the scattering coefficient $f_2<<1$. The scattering coefficient $f_1$ typically ranges from -1 (cell in glycerol \cite{settnes2012forces}) to 1 (glass in water), and is here taken as unity for simplicity. 

\underline{Initial and boundary conditions:} 
We consider an interdigital transducer (IDT), identified by the superscript \(\alpha\) where \(\alpha\) is an element of the set \{L, R, B, T\}. The letters L, R, B, and T correspond to the left, right, bottom, and top boundaries, respectively. In any selected sub-field, denoted as \(j\), the IDT $\alpha$ generates a SAW that is then converted to the pressure wave $p^{\alpha}_j$:

\begin{equation}
    p_j^{\alpha}=p_0 s^{\alpha}_j \sin (\omega_0 t-\varphi^{\alpha}) \Pi(\omega_0 t-\varphi^{\alpha}), \label{eq: excitation}
\end{equation}
where $p_0=\SI{1}{\MPa}$ is the incident wave pressure amplitude, $\omega_0=2\pi f_0$ and  $f_0=\SI{15}{\MHz}$ the excitation frequency. The function $\Pi$ is a gate function that reads $\Pi(\xi)=0$ for all $\xi$ except $\xi\in [ 0,1/2]$. Multiplied by the sine function, this gate function creates a positive half-wave pressure pulse. The relative amplitude coefficient $s^{\alpha}_j$ is independent of the particle position but varies for each combination of sub-fields and transducer and is given in Table~\ref{tab:combi}). The phase offset $\varphi^\alpha$ allows controlling the location $(x_p,y_p)$ of the combined acoustic trap (see Table~\ref{tab:combi}) and depends only on the IDT and trap location. Hence, the sequence of sub-fields to create a trap is independent of the trap location,  up to a delay that is shared by all the sub-fields of a given transducer $\alpha$ and that depends only on the particle location. In a possible experimental realization, one microsequencers or arbitrary function generator per IDT can be used to synthesize successively the four sub-fields required to create a trap at 0-coordinates. The trap could then be moved by using digital delay lines to control the triggering time of each of the microsequencers. Reprogramming the digital delay lines can be done within $ms$ and allow a smooth particle motion.

The total simulation duration $t_{\text{sim}}=0.25\weis$ is chosen to allow the pulse to travel across the channel, and the duration between sub-fields $T/n$ is set to $20\weis$.

\underline{Discretization:} for mesh and solver, the Courant - Friedrichs - Lewy (CFL) number $Co=c_\text{SAW}\Delta t /\Delta x=0.1$, and the minimum number of mesh elements per wavelength is set to $N=12$.

\begin{table}
\caption{\label{tab:combi}Amplitude and phase coefficients of the four sub-fields  }
\begin{ruledtabular}
\begin{tabular}{rrrrrr}
Parameter & $s^{\alpha}_1$ & $s^{\alpha}_2$ & $s^{\alpha}_3$ & $s^{\alpha}_4$ & $\varphi^{\alpha}\footnote{$k=2\pi f_0 / c_d$ is the wavenumber.}$\\
\hline
Left ($L$)  & -1 & $\sqrt{2}$  & 1  & $\sqrt{2}$  & $-k x_p$\\
Right ($R$) & -1 & $-\sqrt{2}$ & 1  & $-\sqrt{2}$ & $k x_p$ \\
Bottom ($B$)&  1 & $-\sqrt{2}$ & -1 & $\sqrt{2}$  & $-k y_p$\\
Top ($T$)   &  1 & $\sqrt{2}$  & -1 & $-\sqrt{2}$ & $k y_p$ \\
\end{tabular}
\end{ruledtabular}
\end{table}

\subsection{Particle motion}
After the ARF field has been computed, it is exported from COMSOL as a \verb|.csv| file, and imported to Python. Our Python script interpolates the COMSOL force field using SciPy \verb|CloughTocher2DInterpolator|. Then Eq. \eqref{eq:v_p} is solved using the Runge Kutta 4-5 (RK45) explicit ordinary differential equation solver with adaptive time stepping (\verb|odeint| in SciPy). When the combined ARF field is updated, the simulation is interrupted by the function ``\verb|update_field_event|'' that loads the newest combined force field.

\section{results \label{results}}
\subsection{Acoustic radiation potential of sub-fields}
\begin{figure*}[!htbp]
    \includegraphics[width=\textwidth]{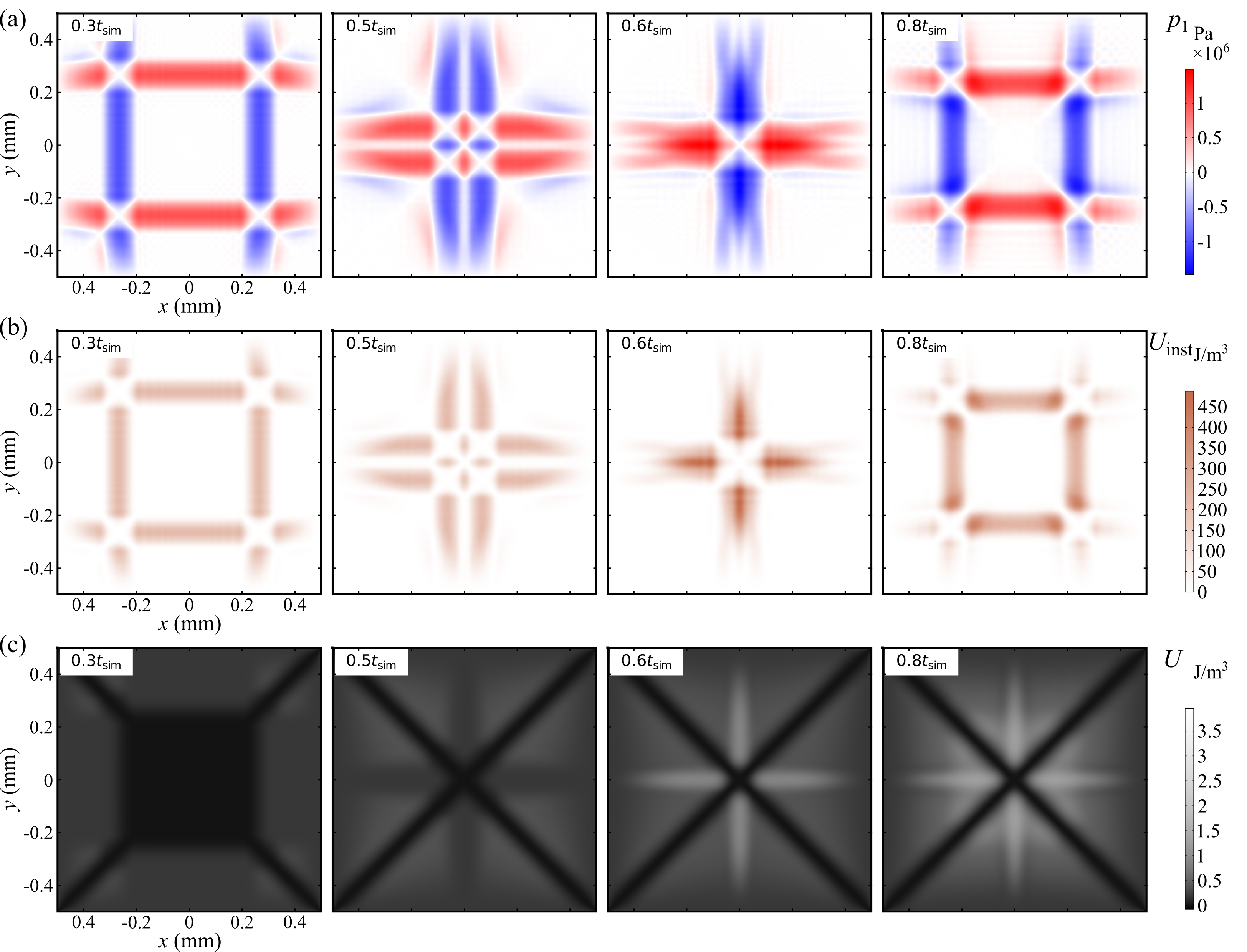}
    \caption{\label{fig:ap}Evolution and gradual build-up of the \textbf{(a)} acoustic pressure \textbf{(b)} instantaneous Gor'kov potential and \textbf{(c)} the actual Gor'kov potential over time for sub-field [1].}
\end{figure*}

We first examine the behavior of acoustic pressure and acoustic radiation potential fields for the first sub-field ($s=s_1$) and setting $(x_p,y_p)=(0,0)$ In Fig.~\ref{fig:ap}(a), we observe the evolution of the acoustic field over time, where pairs of perpendicular pulsed plane waves (blue and red ripples) propagate inward from the domain boundaries. The ripple width depends on the bandwidth of the excitation signal. Destructive interference occurs when waves with opposite phases intersect, such as those from orthogonal upper and lower boundaries, and left and right boundaries. At $t=0.5t_{\text{sim}}$, opposite acoustic waves meet, followed by strong phase interference at $t=0.6t_{\text{sim}}$. Subsequently, the waveform remains constant and propagates towards the opposite side. Notably, the acoustic pressure at the center of the tweezers $(0,0)$ remains zero, creating a constant silent zone.

The instantaneous Gor'kov potential $\mathcal{U}_{1,\text{inst}}$ (Fig.~\ref{fig:ap}(b)) reflects the non-zero acoustic pressure topology. The resulting potential build-up is shown in Fig.~\ref{fig:ap}(c). The potential build-up is nearly completed at $t=0.6t_{\text{sim}}$ although the simulation continues up to $t=t_{\text{sim}}$.  We note that regions experiencing destructive phase interference become potential energy trenches, while those subject to constructive interference morph into potential barriers between the focal point $(0,0)$ and the $x$ and $y$ axes. Hence, the superposition of well-chosen sub-field acoustic radiation potentials can create a singular region of zero potential energy surrounded by a circular barrier, trapping particles exclusively at that focal point.

\begin{figure*}[!htbp]
    \includegraphics[width=\textwidth]{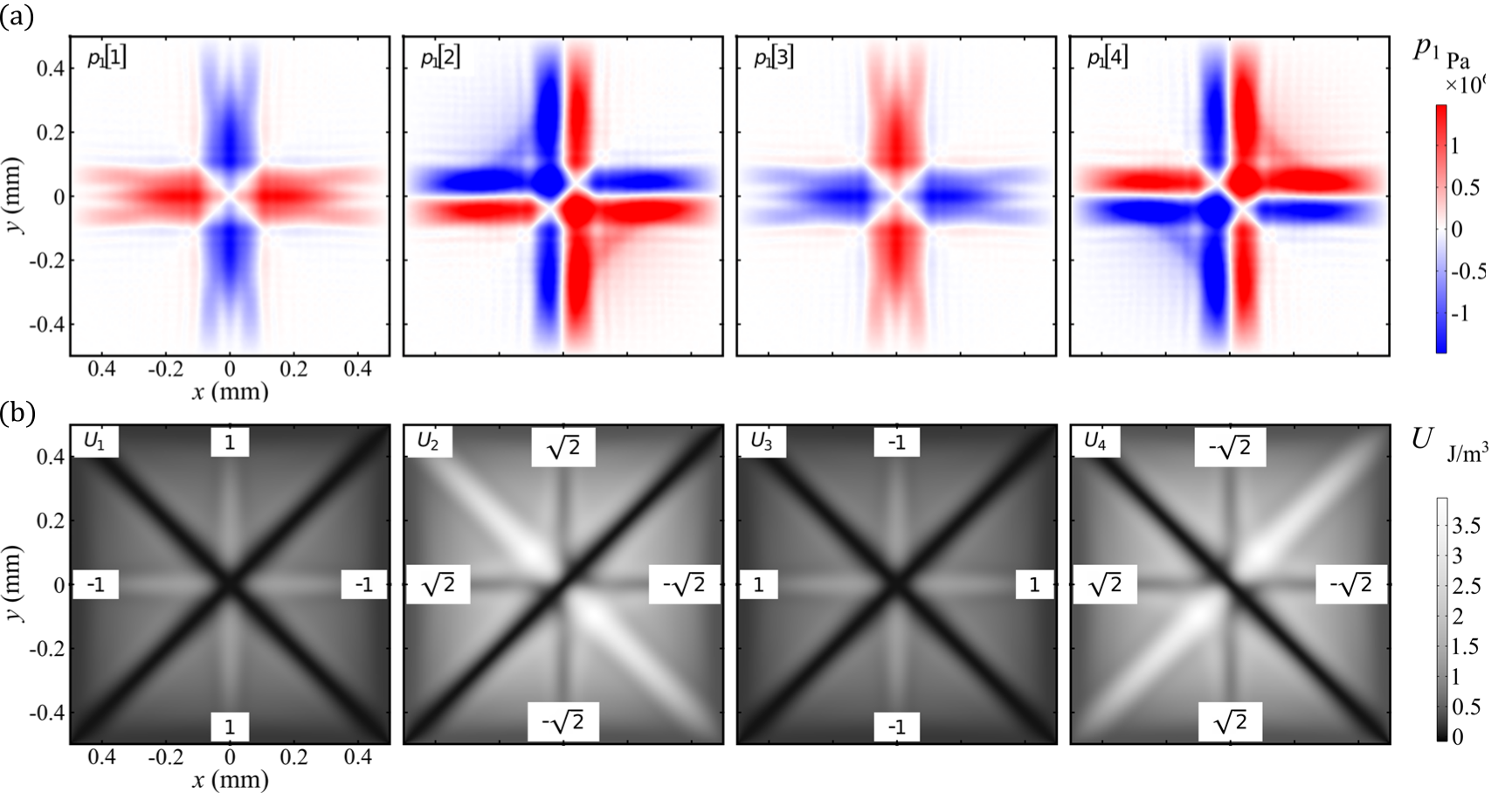}
    \caption{\label{fig:4subfields} snapshot of the acoustic pressure (at $t=0.6t_{\text{sim}}$) and the Gor'kov potential of the four sub-fields used in the study.}
\end{figure*}

Fig.~\ref{fig:4subfields}(a) depicts the acoustic pressure field of the complete set of sub-fields, at a critical moment ($0.6t_{\text{sim}}$) when opposing waves meet (the evolution of the fields is available in SI). Unlike sub-fields [1] and [3], sub-field [2] destructive interference only exists along a \ang{45} diagonal, while sub-field [4] does so along a \ang{135} diagonal. The respective Gor'kov potentials are shown in Fig.~\ref{fig:4subfields}(b). The Gor'kov potential $\mathcal{U}_1$ and $\mathcal{U}_4$ are identical and feature null diagonals. The sub-fields [2] and [4] feature constructive interference along \ang{135} and \ang{45} diagonals, respectively. The key idea when combining sub-fields it to use these constructive interference diagonals to compensate for the destructive interference. We note that since destructive interference only creates null potential in the worst case, one simply has to ensure that the positive potential along the diagonal is equal to the background field. Empirically, we find that setting the amplitude of sub-fields [2] and [4] to $\sqrt{2}$ results in the best balance.

\subsection{Combined acoustic radiation potential}

\begin{figure*}[!htbp]
    \includegraphics[width=0.7\textwidth]{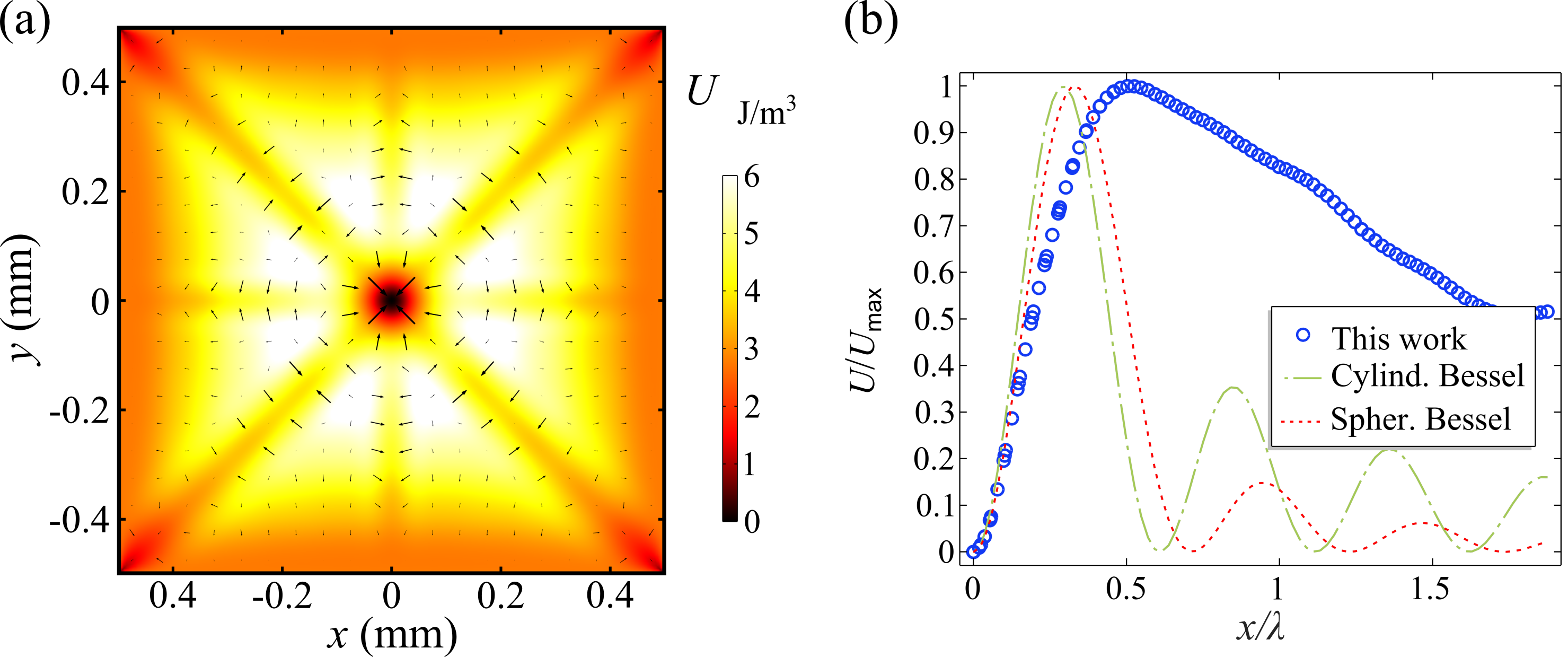}
    \caption{\label{fig:combined_field} \textbf{(a)} Combined Gor'kov potential used in this study. \textbf{(b)} Comparison of the shape of the Gor'kov potential for various acoustic fields.}
\end{figure*}

Fig.~\ref{fig:combined_field}(a) shows the potential distribution of the combined field composed of sub-fields, revealing a clover-leaf pattern with a prominent circular potential well at the center. This addresses the shortcoming of the multi-frequency Fourier synthesis method \cite{xu2021fourier,yang2022harmonic}, where secondary one-dimensional potential trenches were limiting spatial selectivity. Indeed, the new tweezers create a ``pit-trap'', while the vortex tweezers create a trap that looks more like a fenced pen. The pit has a smaller footprint and disrupts less nearby particles, however the pen has the advantage of repelling the nearby particles, which can be advantageous in a crowded environment or when the positioning of other particles is not relevant. 

According to Fig.~\ref{fig:combined_field}(b), the trap remains attractive for a radius of up to half a wavelength approximately, whereas cylindrical and spherical vortex tweezers trapping radius only extends to a quarter wavelength. However, we note that the radius with the maximum trapping force (inflection point of the Gor'kov potential) is similar in all three cases.The maximum force is relevant in the case where the radiation force is balanced by another force, such as friction with the microchannel wall. The potential gradient when moving away from the trapping region is smaller than for the cylindrical trap, which suggests a better selectivity than a cylindrical vortex. This is remarkable because the two-dimensional geometry suggests that it is not possible to focus the energy better than with a Bessel function.

Fig.~\ref{fig:radiation_force_combined_field} shows the ARF of the combined field in the $x$ and $y$ directions. It has a pronounced S-shape, acting as a restoring force on trapped particles. The peak magnitude of the restoring force (\SI{35}{\pico\newton}) exceeds nearly 5 times the background force ($\approx \SI{7}{\pico\newton}$).

\subsection{Particle manipulation}

\begin{figure}[!htbp]
    \includegraphics[width=90mm]{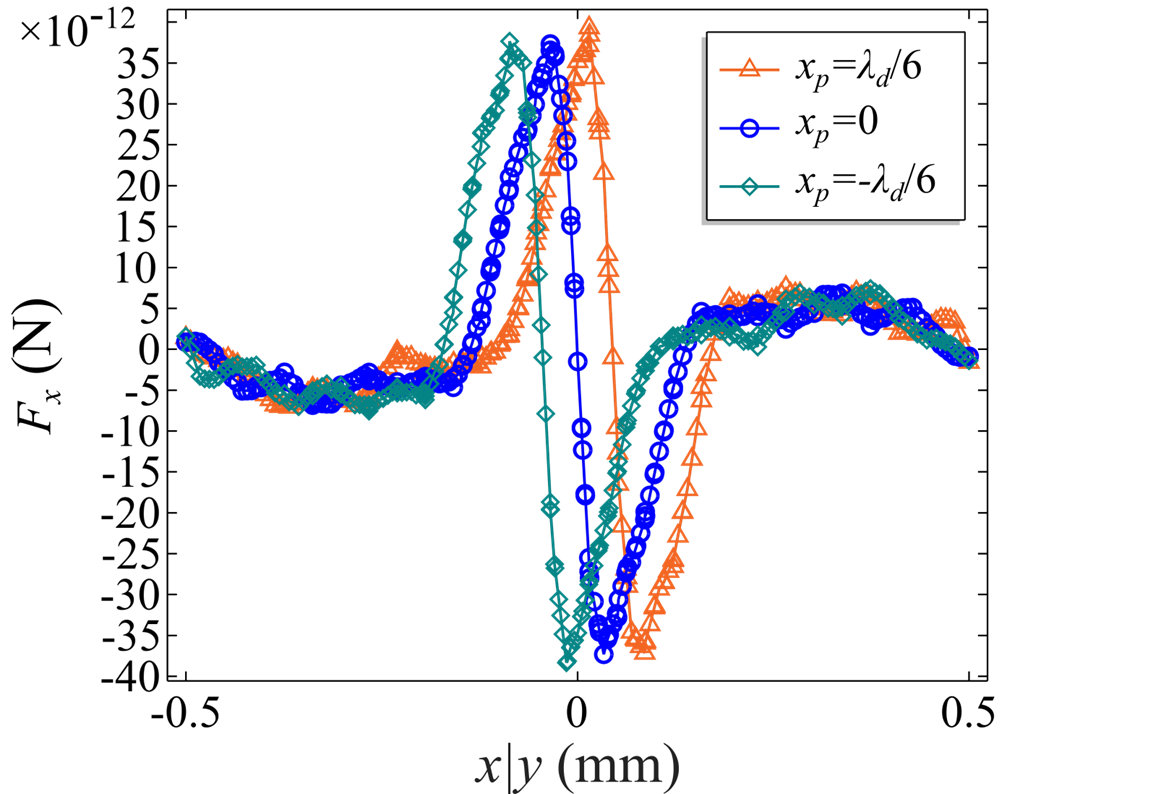}
    \caption{\label{fig:radiation_force_combined_field} Snapshots of the ARF during particle manipulation from  $x_p = - \lambda_d/6$ to $x_p = \lambda_d /6$ by steps of $\lambda_d/6$. The friction force ($F = 22$ pN) requires the successive force fields to intersect above this force threshold in order to maintain particle mobility, and therefore sets a minimum pitch of approximately $\lambda_d/6$.}
\end{figure}
\begin{figure*}[!htbp]
    \includegraphics[width=\textwidth]{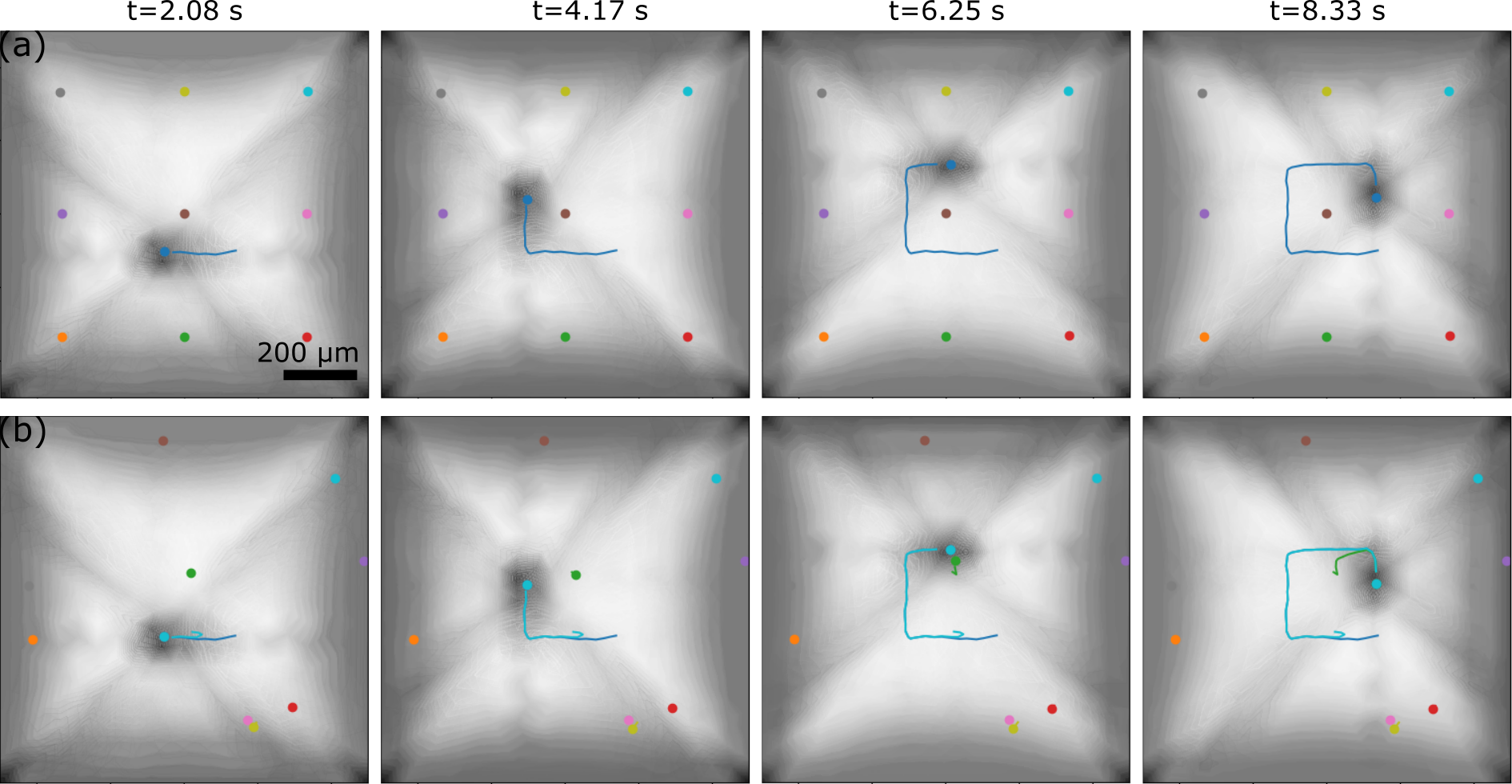}
    \caption{\label{fig:selective_manip} Comet plot of the particle manipulation (the particle instantaneous position is shown as a dot, and the past positions are shown as the ``tail'' of the comet. The shades of gray indicate the combined Gor'kov potential. \textbf{(a)} particles initially positioned in a regular array, with a spacing larger than the trapping radius of the tweezers. \textbf{(b)} particles initially randomly positioned. Some particles are captured by the tweezers during manipulation.} 
\end{figure*}

 The acoustic trap can be moved within the microchannel by adjusting the phase of the $\varphi^\alpha$ in Eq.~\eqref{eq: excitation}, as shown in Fig.~\ref{fig:radiation_force_combined_field}. The time $T$ required to form such potential wells represents the minimum step between subsequent movement processes.
To facilitate the discussion of the ARF distribution, we first restrict the motion of the potential well to the $x$-axis ($y_p=0$) only. We note that the field has a radial symmetry and only the $x$-component of the ARF, $F_x$, needs to be considered. As shown in Fig.~\ref{fig:radiation_force_combined_field}, the ARF profile is translated from $x_p=- \lambda_d /6$ to $+ \lambda_d /6$. The step size of the potential trap movement should be carefully chosen in because the particles can be kept within the trapping at all times only if the potential trap moves by a step smaller than $\lambda_d /6 \approx 44 \weim$. 

This minimal step is set by range of radii where the ARF exceeds the surface friction force (\SI{22}{\pico\newton}). For instance, looking at the transition between $x_p=-\lambda_d/6$ and $x_p=0$, we can see that only particles who experience an ARF larger than the friction force can move. A particle within the trapping range of the tweezers will not fall to the trap center, but instead will move towards it until the friction force balances the ARF ($\mathbf{F}_{rad}=\mathbf{F}_c$), which is the location where  $F_{\mathrm{rad}}=\SI{22}{\pico\newton}$. When the field is updated, the ARF at this location must again exceed the friction threshold. Therefore, one has to ensure that the particle is continuously exposed to \SI{22}{\pico\newton}, which sets the $\lambda_d/6$. We can graphically check that this is the minimal distance because two successive profiles intersect approximately at this force value. In experiments, the friction force is unknown and the step would have to be adjusted depending on the transducer power.

We demonstrate the selective manipulation by simulating a batch of 10 particles dispersed in a microchannel. We aim to move the central particle to describe a square. The manipulation is shown in Fig. \ref{fig:selective_manip}. In the first case (a), all the extra particles are arranged in a regular array away from the trap motion. None of those particles is captured. In the second case (b), the nine other particles are randomly dispersed. While the manipulated particle follows the planned trajectory, the other particles stay nearly in place, demonstrating the selectivity of the manipulation. At \SI{6.25}{\s}, the trap comes too close to an extra particle, which is also captured and then moves along the manipulated particle.

\section{Discussion}
These simulations suggest that can multiplexing could be used to achieve two-dimensional manipulation of particles, addressing the limitation faced by the Fourier tweezers of Xu \etal \cite{xu2021fourier}) or sub time-of-flight tweezers of Collins \etal \cite{collins2016acoustic}. Furthermore, the simulations suggests that this kind of system could also benefit particle array, such as the HANDS platform \cite{yang2022harmonic}, that have been so far limited to affine transformation of the particles coordinates (rotation of the array, scaling, skewing, and translation) and, depending on the stretching, it also allows merging of particles. Indeed, using pulses could allow this type of array tweezers to selectively move one particle independently of all the others.

\section{Conclusion}
Currently, selective manipulation of particles necessitates intricate transducers or transducer arrays. Acoustic manipulation with pulsed waves promises simpler transducers for selective manipulation, but was only demonstrated in one dimension. Our simulations suggest that selective manipulation with pulsed waves could be expanded to two-dimensions by using combination of pulses. Our method leverages the nonlinearity of the Acoustic Radiation Force (ARF) as an intermediary step, enabling the creation of more intricate elementary fields than what could be achieved through mere linear combinations of pressure fields.

This approach could potentially be expanded to incorporate all four transducers, moving beyond the 2+2 combinations utilized in our current work. However, the reason why the weight of $\sqrt{2}$ resulted in optimal selectivity is still unclear and warrants further investigation.

The next significant step towards achieving two-dimensional selective manipulation with straightforward acoustic tweezers would be an experimental demonstration of this type of device. These newly developed tweezers, however, impose strict constraints on the transducer bandwidth. It remains to be determined whether using pulses, as opposed to chirps, is the most effective method for manipulation, especially considering the potential risk of dielectric breakdown of the interdigital transducer (IDT).

As we look to the future, we anticipate that continued advancements in the field of pulsed manipulation could spur the development of wide bandwidth transducers, specifically designed for micro-acoustofluidics.

\begin{acknowledgments}
This work was supported by the National Natural Science Foundation of China with Grant No. 62274039.	
\end{acknowledgments}


\bibliography{2DSAT-sim-ref}

\begin{thebibliography}{22}%
\makeatletter
\providecommand \@ifxundefined [1]{%
 \@ifx{#1\undefined}
}%
\providecommand \@ifnum [1]{%
 \ifnum #1\expandafter \@firstoftwo
 \else \expandafter \@secondoftwo
 \fi
}%
\providecommand \@ifx [1]{%
 \ifx #1\expandafter \@firstoftwo
 \else \expandafter \@secondoftwo
 \fi
}%
\providecommand \natexlab [1]{#1}%
\providecommand \enquote  [1]{``#1''}%
\providecommand \bibnamefont  [1]{#1}%
\providecommand \bibfnamefont [1]{#1}%
\providecommand \citenamefont [1]{#1}%
\providecommand \href@noop [0]{\@secondoftwo}%
\providecommand \href [0]{\begingroup \@sanitize@url \@href}%
\providecommand \@href[1]{\@@startlink{#1}\@@href}%
\providecommand \@@href[1]{\endgroup#1\@@endlink}%
\providecommand \@sanitize@url [0]{\catcode `\\12\catcode `\$12\catcode `\&12\catcode `\#12\catcode `\^12\catcode `\_12\catcode `\%12\relax}%
\providecommand \@@startlink[1]{}%
\providecommand \@@endlink[0]{}%
\providecommand \url  [0]{\begingroup\@sanitize@url \@url }%
\providecommand \@url [1]{\endgroup\@href {#1}{\urlprefix }}%
\providecommand \urlprefix  [0]{URL }%
\providecommand \Eprint [0]{\href }%
\providecommand \doibase [0]{https://doi.org/}%
\providecommand \selectlanguage [0]{\@gobble}%
\providecommand \bibinfo  [0]{\@secondoftwo}%
\providecommand \bibfield  [0]{\@secondoftwo}%
\providecommand \translation [1]{[#1]}%
\providecommand \BibitemOpen [0]{}%
\providecommand \bibitemStop [0]{}%
\providecommand \bibitemNoStop [0]{.\EOS\space}%
\providecommand \EOS [0]{\spacefactor3000\relax}%
\providecommand \BibitemShut  [1]{\csname bibitem#1\endcsname}%
\let\auto@bib@innerbib\@empty
\bibitem [{\citenamefont {Thomas}\ \emph {et~al.}(2017)\citenamefont {Thomas}, \citenamefont {Marchiano},\ and\ \citenamefont {Baresch}}]{thomas2017acoustical}%
  \BibitemOpen
  \bibfield  {author} {\bibinfo {author} {\bibfnamefont {J.-L.}\ \bibnamefont {Thomas}}, \bibinfo {author} {\bibfnamefont {R.}~\bibnamefont {Marchiano}},\ and\ \bibinfo {author} {\bibfnamefont {D.}~\bibnamefont {Baresch}},\ }\bibfield  {title} {\bibinfo {title} {Acoustical and optical radiation pressure and the development of single beam acoustical tweezers},\ }\href {https://doi.org/10.1016/j.jqsrt.2017.01.012} {\bibfield  {journal} {\bibinfo  {journal} {J. Quant. Spectrosc. Ra.}\ }\textbf {\bibinfo {volume} {195}},\ \bibinfo {pages} {55} (\bibinfo {year} {2017})}\BibitemShut {NoStop}%
\bibitem [{\citenamefont {Cui}\ \emph {et~al.}(2023)\citenamefont {Cui}, \citenamefont {Dutcher}, \citenamefont {Bayly},\ and\ \citenamefont {Meacham}}]{cui2023robust}%
  \BibitemOpen
  \bibfield  {author} {\bibinfo {author} {\bibfnamefont {M.}~\bibnamefont {Cui}}, \bibinfo {author} {\bibfnamefont {S.~K.}\ \bibnamefont {Dutcher}}, \bibinfo {author} {\bibfnamefont {P.~V.}\ \bibnamefont {Bayly}},\ and\ \bibinfo {author} {\bibfnamefont {J.~M.}\ \bibnamefont {Meacham}},\ }\bibfield  {title} {\bibinfo {title} {Robust acoustic trapping and perturbation of single-cell microswimmers illuminate three-dimensional swimming and ciliary coordination},\ }\href {https://doi.org/10.1073/pnas.2218951120} {\bibfield  {journal} {\bibinfo  {journal} {Proc. Natl. Acad. Sci.}\ }\textbf {\bibinfo {volume} {120}},\ \bibinfo {pages} {e2218951120} (\bibinfo {year} {2023})}\BibitemShut {NoStop}%
\bibitem [{\citenamefont {Guo}\ \emph {et~al.}(2014)\citenamefont {Guo}, \citenamefont {Li}, \citenamefont {French}, \citenamefont {Mao}, \citenamefont {Zhao}, \citenamefont {Li}, \citenamefont {Nama}, \citenamefont {Fick}, \citenamefont {Benkovic},\ and\ \citenamefont {Huang}}]{guo2014controlling}%
  \BibitemOpen
  \bibfield  {author} {\bibinfo {author} {\bibfnamefont {F.}~\bibnamefont {Guo}}, \bibinfo {author} {\bibfnamefont {P.}~\bibnamefont {Li}}, \bibinfo {author} {\bibfnamefont {J.~B.}\ \bibnamefont {French}}, \bibinfo {author} {\bibfnamefont {Z.}~\bibnamefont {Mao}}, \bibinfo {author} {\bibfnamefont {H.}~\bibnamefont {Zhao}}, \bibinfo {author} {\bibfnamefont {S.}~\bibnamefont {Li}}, \bibinfo {author} {\bibfnamefont {N.}~\bibnamefont {Nama}}, \bibinfo {author} {\bibfnamefont {J.~R.}\ \bibnamefont {Fick}}, \bibinfo {author} {\bibfnamefont {S.~J.}\ \bibnamefont {Benkovic}},\ and\ \bibinfo {author} {\bibfnamefont {T.~J.}\ \bibnamefont {Huang}},\ }\bibfield  {title} {\bibinfo {title} {Controlling cell\textendash cell interactions using surface acoustic waves},\ }\href {https://doi.org/10.1073/pnas.1422068112} {\bibfield  {journal} {\bibinfo  {journal} {Proc. Natl. Acad. Sci.}\ }\textbf {\bibinfo {volume} {112}},\ \bibinfo {pages} {43} (\bibinfo {year} {2014})}\BibitemShut {NoStop}%
\bibitem [{\citenamefont {Yang}\ \emph {et~al.}(2022)\citenamefont {Yang}, \citenamefont {Tian}, \citenamefont {Wang}, \citenamefont {Rufo}, \citenamefont {Li}, \citenamefont {Mai}, \citenamefont {Xia}, \citenamefont {Bachman}, \citenamefont {Huang}, \citenamefont {Wu}, \citenamefont {Chen}, \citenamefont {Lee},\ and\ \citenamefont {Huang}}]{yang2022harmonic}%
  \BibitemOpen
  \bibfield  {author} {\bibinfo {author} {\bibfnamefont {S.}~\bibnamefont {Yang}}, \bibinfo {author} {\bibfnamefont {Z.}~\bibnamefont {Tian}}, \bibinfo {author} {\bibfnamefont {Z.}~\bibnamefont {Wang}}, \bibinfo {author} {\bibfnamefont {J.}~\bibnamefont {Rufo}}, \bibinfo {author} {\bibfnamefont {P.}~\bibnamefont {Li}}, \bibinfo {author} {\bibfnamefont {J.}~\bibnamefont {Mai}}, \bibinfo {author} {\bibfnamefont {J.}~\bibnamefont {Xia}}, \bibinfo {author} {\bibfnamefont {H.}~\bibnamefont {Bachman}}, \bibinfo {author} {\bibfnamefont {P.-H.}\ \bibnamefont {Huang}}, \bibinfo {author} {\bibfnamefont {M.}~\bibnamefont {Wu}}, \bibinfo {author} {\bibfnamefont {C.}~\bibnamefont {Chen}}, \bibinfo {author} {\bibfnamefont {L.~P.}\ \bibnamefont {Lee}},\ and\ \bibinfo {author} {\bibfnamefont {T.~J.}\ \bibnamefont {Huang}},\ }\bibfield  {title} {\bibinfo {title} {Harmonic acoustics for dynamic and selective particle manipulation},\ }\href {https://doi.org/10.1038/s41563-022-01210-8} {\bibfield  {journal} {\bibinfo  {journal}
  {Nat. Mater.}\ }\textbf {\bibinfo {volume} {21}},\ \bibinfo {pages} {540} (\bibinfo {year} {2022})}\BibitemShut {NoStop}%
\bibitem [{\citenamefont {Baudoin}\ \emph {et~al.}(2019)\citenamefont {Baudoin}, \citenamefont {Gerbedoen}, \citenamefont {Riaud}, \citenamefont {Matar}, \citenamefont {Smagin},\ and\ \citenamefont {Thomas}}]{baudoin2019folding}%
  \BibitemOpen
  \bibfield  {author} {\bibinfo {author} {\bibfnamefont {M.}~\bibnamefont {Baudoin}}, \bibinfo {author} {\bibfnamefont {J.-C.}\ \bibnamefont {Gerbedoen}}, \bibinfo {author} {\bibfnamefont {A.}~\bibnamefont {Riaud}}, \bibinfo {author} {\bibfnamefont {O.~B.}\ \bibnamefont {Matar}}, \bibinfo {author} {\bibfnamefont {N.}~\bibnamefont {Smagin}},\ and\ \bibinfo {author} {\bibfnamefont {J.-L.}\ \bibnamefont {Thomas}},\ }\bibfield  {title} {\bibinfo {title} {Folding a focalized acoustical vortex on a flat holographic transducer: Miniaturized selective acoustical tweezers},\ }\href {https://doi.org/10.1126/sciadv.aav1967} {\bibfield  {journal} {\bibinfo  {journal} {Sci. Adv.}\ }\textbf {\bibinfo {volume} {5}},\ \bibinfo {pages} {eaav1967} (\bibinfo {year} {2019})}\BibitemShut {NoStop}%
\bibitem [{\citenamefont {Baudoin}\ \emph {et~al.}(2020)\citenamefont {Baudoin}, \citenamefont {Thomas}, \citenamefont {Sahely}, \citenamefont {Gerbedoen}, \citenamefont {Gong}, \citenamefont {Sivery}, \citenamefont {Matar}, \citenamefont {Smagin}, \citenamefont {Favreau},\ and\ \citenamefont {Vlandas}}]{baudoin2020spatially}%
  \BibitemOpen
  \bibfield  {author} {\bibinfo {author} {\bibfnamefont {M.}~\bibnamefont {Baudoin}}, \bibinfo {author} {\bibfnamefont {J.-L.}\ \bibnamefont {Thomas}}, \bibinfo {author} {\bibfnamefont {R.~A.}\ \bibnamefont {Sahely}}, \bibinfo {author} {\bibfnamefont {J.-C.}\ \bibnamefont {Gerbedoen}}, \bibinfo {author} {\bibfnamefont {Z.}~\bibnamefont {Gong}}, \bibinfo {author} {\bibfnamefont {A.}~\bibnamefont {Sivery}}, \bibinfo {author} {\bibfnamefont {O.~B.}\ \bibnamefont {Matar}}, \bibinfo {author} {\bibfnamefont {N.}~\bibnamefont {Smagin}}, \bibinfo {author} {\bibfnamefont {P.}~\bibnamefont {Favreau}},\ and\ \bibinfo {author} {\bibfnamefont {A.}~\bibnamefont {Vlandas}},\ }\bibfield  {title} {\bibinfo {title} {{Spatially selective manipulation of cells with single-beam acoustical tweezers}},\ }\href {https://doi.org/10.1038/s41467-020-18000-y} {\bibfield  {journal} {\bibinfo  {journal} {Nat. Commun.}\ }\textbf {\bibinfo {volume} {11}},\ \bibinfo {pages} {4244} (\bibinfo {year} {2020})}\BibitemShut {NoStop}%
\bibitem [{\citenamefont {Royer}\ and\ \citenamefont {Dieulesaint}(1999)}]{royer1999elastic}%
  \BibitemOpen
  \bibfield  {author} {\bibinfo {author} {\bibfnamefont {D.}~\bibnamefont {Royer}}\ and\ \bibinfo {author} {\bibfnamefont {E.}~\bibnamefont {Dieulesaint}},\ }\href@noop {} {\emph {\bibinfo {title} {Elastic Waves in Solids II: Generation, Acousto-optic Interaction, Applications}}}\ (\bibinfo  {publisher} {Springer Science \& Business Media},\ \bibinfo {year} {1999})\BibitemShut {NoStop}%
\bibitem [{\citenamefont {Williamson}\ and\ \citenamefont {Smith}(1973)}]{williamson1973use}%
  \BibitemOpen
  \bibfield  {author} {\bibinfo {author} {\bibfnamefont {R.~C.}\ \bibnamefont {Williamson}}\ and\ \bibinfo {author} {\bibfnamefont {H.~I.}\ \bibnamefont {Smith}},\ }\bibfield  {title} {\bibinfo {title} {{The Use of Surface-Elastic-Wave Reflection Gratings in Large Time-Bandwidth Pulse-Compression Filters}},\ }\href {https://doi.org/10.1109/tmtt.1973.1127970} {\bibfield  {journal} {\bibinfo  {journal} {IEEE Trans. Microwave Theory Tech.}\ }\textbf {\bibinfo {volume} {21}},\ \bibinfo {pages} {195} (\bibinfo {year} {1973})}\BibitemShut {NoStop}%
\bibitem [{\citenamefont {Wang}\ \emph {et~al.}(2022{\natexlab{a}})\citenamefont {Wang}, \citenamefont {Ota}, \citenamefont {Edlbauer}, \citenamefont {Jadot}, \citenamefont {Mortemousque}, \citenamefont {Richard}, \citenamefont {Okazaki}, \citenamefont {Nakamura}, \citenamefont {Ludwig}, \citenamefont {Wieck}, \citenamefont {Urdampilleta}, \citenamefont {Meunier}, \citenamefont {Kodera}, \citenamefont {Kaneko}, \citenamefont {Takada},\ and\ \citenamefont {B{\"a}uerle}}]{wang2022generation}%
  \BibitemOpen
  \bibfield  {author} {\bibinfo {author} {\bibfnamefont {J.}~\bibnamefont {Wang}}, \bibinfo {author} {\bibfnamefont {S.}~\bibnamefont {Ota}}, \bibinfo {author} {\bibfnamefont {H.}~\bibnamefont {Edlbauer}}, \bibinfo {author} {\bibfnamefont {B.}~\bibnamefont {Jadot}}, \bibinfo {author} {\bibfnamefont {P.-A.}\ \bibnamefont {Mortemousque}}, \bibinfo {author} {\bibfnamefont {A.}~\bibnamefont {Richard}}, \bibinfo {author} {\bibfnamefont {Y.}~\bibnamefont {Okazaki}}, \bibinfo {author} {\bibfnamefont {S.}~\bibnamefont {Nakamura}}, \bibinfo {author} {\bibfnamefont {A.}~\bibnamefont {Ludwig}}, \bibinfo {author} {\bibfnamefont {A.~D.}\ \bibnamefont {Wieck}}, \bibinfo {author} {\bibfnamefont {M.}~\bibnamefont {Urdampilleta}}, \bibinfo {author} {\bibfnamefont {T.}~\bibnamefont {Meunier}}, \bibinfo {author} {\bibfnamefont {T.}~\bibnamefont {Kodera}}, \bibinfo {author} {\bibfnamefont {N.-H.}\ \bibnamefont {Kaneko}}, \bibinfo {author} {\bibfnamefont {S.}~\bibnamefont {Takada}},\ and\ \bibinfo {author} {\bibfnamefont
  {C.}~\bibnamefont {B{\"a}uerle}},\ }\bibfield  {title} {\bibinfo {title} {{Generation of a Single-Cycle Acoustic Pulse: A Scalable Solution for Transport in Single-Electron Circuits}},\ }\href {https://doi.org/10.1103/physrevx.12.031035} {\bibfield  {journal} {\bibinfo  {journal} {Phys. Rev. X}\ }\textbf {\bibinfo {volume} {12}},\ \bibinfo {pages} {031035} (\bibinfo {year} {2022}{\natexlab{a}})}\BibitemShut {NoStop}%
\bibitem [{\citenamefont {Song}\ \emph {et~al.}(2022)\citenamefont {Song}, \citenamefont {Wang}, \citenamefont {Zhou},\ and\ \citenamefont {Riaud}}]{song2022design}%
  \BibitemOpen
  \bibfield  {author} {\bibinfo {author} {\bibfnamefont {S.}~\bibnamefont {Song}}, \bibinfo {author} {\bibfnamefont {Q.}~\bibnamefont {Wang}}, \bibinfo {author} {\bibfnamefont {J.}~\bibnamefont {Zhou}},\ and\ \bibinfo {author} {\bibfnamefont {A.}~\bibnamefont {Riaud}},\ }\bibfield  {title} {\bibinfo {title} {Design of interdigitated transducers for acoustofluidic applications},\ }\href {https://doi.org/10.1063/10.0013405} {\bibfield  {journal} {\bibinfo  {journal} {Nanotechnology and Precision Engineering}\ }\textbf {\bibinfo {volume} {5}},\ \bibinfo {pages} {035001} (\bibinfo {year} {2022})}\BibitemShut {NoStop}%
\bibitem [{\citenamefont {Li}\ \emph {et~al.}(2021)\citenamefont {Li}, \citenamefont {Zhong}, \citenamefont {Liu}, \citenamefont {Lu}, \citenamefont {Liang},\ and\ \citenamefont {Ai}}]{li2021physical}%
  \BibitemOpen
  \bibfield  {author} {\bibinfo {author} {\bibfnamefont {P.}~\bibnamefont {Li}}, \bibinfo {author} {\bibfnamefont {J.}~\bibnamefont {Zhong}}, \bibinfo {author} {\bibfnamefont {N.}~\bibnamefont {Liu}}, \bibinfo {author} {\bibfnamefont {X.}~\bibnamefont {Lu}}, \bibinfo {author} {\bibfnamefont {M.}~\bibnamefont {Liang}},\ and\ \bibinfo {author} {\bibfnamefont {Y.}~\bibnamefont {Ai}},\ }\bibfield  {title} {\bibinfo {title} {Physical properties-based microparticle sorting at submicron resolution using a tunable acoustofluidic device},\ }\href {https://doi.org/10.1016/j.snb.2021.130203} {\bibfield  {journal} {\bibinfo  {journal} {Sens. Actuators B Chem.}\ }\textbf {\bibinfo {volume} {344}},\ \bibinfo {pages} {130203} (\bibinfo {year} {2021})}\BibitemShut {NoStop}%
\bibitem [{\citenamefont {Schmid}\ \emph {et~al.}(2014)\citenamefont {Schmid}, \citenamefont {Weitz},\ and\ \citenamefont {Franke}}]{schmid2014sorting}%
  \BibitemOpen
  \bibfield  {author} {\bibinfo {author} {\bibfnamefont {L.}~\bibnamefont {Schmid}}, \bibinfo {author} {\bibfnamefont {D.~A.}\ \bibnamefont {Weitz}},\ and\ \bibinfo {author} {\bibfnamefont {T.}~\bibnamefont {Franke}},\ }\bibfield  {title} {\bibinfo {title} {Sorting drops and cells with acoustics: acoustic microfluidic fluorescence-activated cell sorter},\ }\href {https://doi.org/10.1039/C4LC00588K} {\bibfield  {journal} {\bibinfo  {journal} {Lab Chip}\ }\textbf {\bibinfo {volume} {14}},\ \bibinfo {pages} {3710} (\bibinfo {year} {2014})}\BibitemShut {NoStop}%
\bibitem [{\citenamefont {Khalid}\ \emph {et~al.}(2013)\citenamefont {Khalid}, \citenamefont {Reboud}, \citenamefont {Wilson},\ and\ \citenamefont {Cooper}}]{khalid2013control}%
  \BibitemOpen
  \bibfield  {author} {\bibinfo {author} {\bibfnamefont {M.~A.}\ \bibnamefont {Khalid}}, \bibinfo {author} {\bibfnamefont {J.}~\bibnamefont {Reboud}}, \bibinfo {author} {\bibfnamefont {R.}~\bibnamefont {Wilson}},\ and\ \bibinfo {author} {\bibfnamefont {J.~M.}\ \bibnamefont {Cooper}},\ }\bibfield  {title} {\bibinfo {title} {Control of blood rheological properties using surface acoustic waves},\ }in\ \href@noop {} {\emph {\bibinfo {booktitle} {17th International Conference on Miniaturized Systems for Chemistry and Life Sciences}}}\ (\bibinfo {year} {2013})\ pp.\ \bibinfo {pages} {101--103}\BibitemShut {NoStop}%
\bibitem [{\citenamefont {Collins}\ \emph {et~al.}(2016)\citenamefont {Collins}, \citenamefont {Devendran}, \citenamefont {Ma}, \citenamefont {Ng}, \citenamefont {Neild},\ and\ \citenamefont {Ai}}]{collins2016acoustic}%
  \BibitemOpen
  \bibfield  {author} {\bibinfo {author} {\bibfnamefont {D.~J.}\ \bibnamefont {Collins}}, \bibinfo {author} {\bibfnamefont {C.}~\bibnamefont {Devendran}}, \bibinfo {author} {\bibfnamefont {Z.}~\bibnamefont {Ma}}, \bibinfo {author} {\bibfnamefont {J.~W.}\ \bibnamefont {Ng}}, \bibinfo {author} {\bibfnamefont {A.}~\bibnamefont {Neild}},\ and\ \bibinfo {author} {\bibfnamefont {Y.}~\bibnamefont {Ai}},\ }\bibfield  {title} {\bibinfo {title} {Acoustic tweezers via sub\textendash time-of-flight regime surface acoustic waves},\ }\href {https://doi.org/10.1126/sciadv.1600089} {\bibfield  {journal} {\bibinfo  {journal} {Sci. Adv.}\ }\textbf {\bibinfo {volume} {2}},\ \bibinfo {pages} {e1600089} (\bibinfo {year} {2016})}\BibitemShut {NoStop}%
\bibitem [{\citenamefont {Xu}\ \emph {et~al.}(2021)\citenamefont {Xu}, \citenamefont {Huang}, \citenamefont {Zhang}, \citenamefont {Xie}, \citenamefont {Ni}, \citenamefont {Huang}, \citenamefont {Yao}, \citenamefont {Tu}, \citenamefont {Guo},\ and\ \citenamefont {Zhang}}]{xu2021fourier}%
  \BibitemOpen
  \bibfield  {author} {\bibinfo {author} {\bibfnamefont {G.}~\bibnamefont {Xu}}, \bibinfo {author} {\bibfnamefont {J.}~\bibnamefont {Huang}}, \bibinfo {author} {\bibfnamefont {Y.}~\bibnamefont {Zhang}}, \bibinfo {author} {\bibfnamefont {L.}~\bibnamefont {Xie}}, \bibinfo {author} {\bibfnamefont {Z.}~\bibnamefont {Ni}}, \bibinfo {author} {\bibfnamefont {C.}~\bibnamefont {Huang}}, \bibinfo {author} {\bibfnamefont {G.}~\bibnamefont {Yao}}, \bibinfo {author} {\bibfnamefont {J.}~\bibnamefont {Tu}}, \bibinfo {author} {\bibfnamefont {X.}~\bibnamefont {Guo}},\ and\ \bibinfo {author} {\bibfnamefont {D.}~\bibnamefont {Zhang}},\ }\bibfield  {title} {\bibinfo {title} {{Fourier Acoustical Tweezers: Synthesizing Arbitrary Radiation Force Using Nonmonochromatic Waves on Discrete-Frequency Basis}},\ }\href {https://doi.org/10.1103/physrevapplied.15.044037} {\bibfield  {journal} {\bibinfo  {journal} {Phys. Rev. Appl.}\ }\textbf {\bibinfo {volume} {15}},\ \bibinfo {pages} {044037} (\bibinfo {year} {2021})}\BibitemShut {NoStop}%
\bibitem [{\citenamefont {Wang}\ \emph {et~al.}(2021)\citenamefont {Wang}, \citenamefont {Riaud}, \citenamefont {Zhou}, \citenamefont {Gong},\ and\ \citenamefont {Baudoin}}]{wang2021acoustic}%
  \BibitemOpen
  \bibfield  {author} {\bibinfo {author} {\bibfnamefont {Q.}~\bibnamefont {Wang}}, \bibinfo {author} {\bibfnamefont {A.}~\bibnamefont {Riaud}}, \bibinfo {author} {\bibfnamefont {J.}~\bibnamefont {Zhou}}, \bibinfo {author} {\bibfnamefont {Z.}~\bibnamefont {Gong}},\ and\ \bibinfo {author} {\bibfnamefont {M.}~\bibnamefont {Baudoin}},\ }\bibfield  {title} {\bibinfo {title} {{Acoustic Radiation Force on Small Spheres Due to Transient Acoustic Fields}},\ }\href {https://doi.org/10.1103/PhysRevApplied.15.044034} {\bibfield  {journal} {\bibinfo  {journal} {Phys. Rev. Appl.}\ }\textbf {\bibinfo {volume} {15}},\ \bibinfo {pages} {044034} (\bibinfo {year} {2021})}\BibitemShut {NoStop}%
\bibitem [{\citenamefont {Chen}\ \emph {et~al.}(2023)\citenamefont {Chen}, \citenamefont {Wang}, \citenamefont {Wang}, \citenamefont {Zhou},\ and\ \citenamefont {Riaud}}]{chen2023numerical}%
  \BibitemOpen
  \bibfield  {author} {\bibinfo {author} {\bibfnamefont {S.}~\bibnamefont {Chen}}, \bibinfo {author} {\bibfnamefont {Q.}~\bibnamefont {Wang}}, \bibinfo {author} {\bibfnamefont {Q.}~\bibnamefont {Wang}}, \bibinfo {author} {\bibfnamefont {J.}~\bibnamefont {Zhou}},\ and\ \bibinfo {author} {\bibfnamefont {A.}~\bibnamefont {Riaud}},\ }\bibfield  {title} {\bibinfo {title} {{Numerical Simulation of the Radiation Force from Transient Acoustic Fields: Application to Laser-Guided Acoustic Tweezers}},\ }\href {https://doi.org/10.1103/PhysRevApplied.19.054057} {\bibfield  {journal} {\bibinfo  {journal} {Phys. Rev. Appl.}\ }\textbf {\bibinfo {volume} {19}},\ \bibinfo {pages} {054057} (\bibinfo {year} {2023})}\BibitemShut {NoStop}%
\bibitem [{\citenamefont {Wang}\ \emph {et~al.}(2022{\natexlab{b}})\citenamefont {Wang}, \citenamefont {Chen}, \citenamefont {Zhou},\ and\ \citenamefont {Riaud}}]{wang2022laser}%
  \BibitemOpen
  \bibfield  {author} {\bibinfo {author} {\bibfnamefont {Q.}~\bibnamefont {Wang}}, \bibinfo {author} {\bibfnamefont {S.}~\bibnamefont {Chen}}, \bibinfo {author} {\bibfnamefont {J.}~\bibnamefont {Zhou}},\ and\ \bibinfo {author} {\bibfnamefont {A.}~\bibnamefont {Riaud}},\ }\bibfield  {title} {\bibinfo {title} {{Laser-guided acoustic tweezers}},\ }\Eprint {https://arxiv.org/abs/2203.14497} {arXiv:2203.14497}  (\bibinfo {year} {2022}{\natexlab{b}})\BibitemShut {NoStop}%
\bibitem [{\citenamefont {Bernard}\ \emph {et~al.}(2017)\citenamefont {Bernard}, \citenamefont {Doinikov}, \citenamefont {Marmottant}, \citenamefont {Rabaud}, \citenamefont {Poulain},\ and\ \citenamefont {Thibault}}]{bernard2017controlled}%
  \BibitemOpen
  \bibfield  {author} {\bibinfo {author} {\bibfnamefont {I.}~\bibnamefont {Bernard}}, \bibinfo {author} {\bibfnamefont {A.~A.}\ \bibnamefont {Doinikov}}, \bibinfo {author} {\bibfnamefont {P.}~\bibnamefont {Marmottant}}, \bibinfo {author} {\bibfnamefont {D.}~\bibnamefont {Rabaud}}, \bibinfo {author} {\bibfnamefont {C.}~\bibnamefont {Poulain}},\ and\ \bibinfo {author} {\bibfnamefont {P.}~\bibnamefont {Thibault}},\ }\bibfield  {title} {\bibinfo {title} {Controlled rotation and translation of spherical particles or living cells by surface acoustic waves},\ }\href {https://doi.org/10.1039/c7lc00084g} {\bibfield  {journal} {\bibinfo  {journal} {Lab Chip}\ }\textbf {\bibinfo {volume} {17}},\ \bibinfo {pages} {2470} (\bibinfo {year} {2017})}\BibitemShut {NoStop}%
\bibitem [{\citenamefont {Riaud}\ \emph {et~al.}(2015)\citenamefont {Riaud}, \citenamefont {Thomas}, \citenamefont {Baudoin},\ and\ \citenamefont {Bou~Matar}}]{riaud2015taming}%
  \BibitemOpen
  \bibfield  {author} {\bibinfo {author} {\bibfnamefont {A.}~\bibnamefont {Riaud}}, \bibinfo {author} {\bibfnamefont {J.-L.}\ \bibnamefont {Thomas}}, \bibinfo {author} {\bibfnamefont {M.}~\bibnamefont {Baudoin}},\ and\ \bibinfo {author} {\bibfnamefont {O.}~\bibnamefont {Bou~Matar}},\ }\bibfield  {title} {\bibinfo {title} {{Taming the degeneration of Bessel beams at an anisotropic-isotropic interface: Toward three-dimensional control of confined vortical waves}},\ }\href {https://doi.org/10.1103/physreve.92.063201} {\bibfield  {journal} {\bibinfo  {journal} {Phys. Rev. E}\ }\textbf {\bibinfo {volume} {92}},\ \bibinfo {pages} {063201} (\bibinfo {year} {2015})}\BibitemShut {NoStop}%
\bibitem [{\citenamefont {Simon}\ \emph {et~al.}(2017)\citenamefont {Simon}, \citenamefont {Andrade}, \citenamefont {Reboud}, \citenamefont {Marques-Hueso}, \citenamefont {Desmulliez}, \citenamefont {Cooper}, \citenamefont {Riehle},\ and\ \citenamefont {Bernassau}}]{simon2017particle}%
  \BibitemOpen
  \bibfield  {author} {\bibinfo {author} {\bibfnamefont {G.}~\bibnamefont {Simon}}, \bibinfo {author} {\bibfnamefont {M.~A.~B.}\ \bibnamefont {Andrade}}, \bibinfo {author} {\bibfnamefont {J.}~\bibnamefont {Reboud}}, \bibinfo {author} {\bibfnamefont {J.}~\bibnamefont {Marques-Hueso}}, \bibinfo {author} {\bibfnamefont {M.~P.~Y.}\ \bibnamefont {Desmulliez}}, \bibinfo {author} {\bibfnamefont {J.~M.}\ \bibnamefont {Cooper}}, \bibinfo {author} {\bibfnamefont {M.~O.}\ \bibnamefont {Riehle}},\ and\ \bibinfo {author} {\bibfnamefont {A.~L.}\ \bibnamefont {Bernassau}},\ }\bibfield  {title} {\bibinfo {title} {Particle separation by phase modulated surface acoustic waves},\ }\href {https://doi.org/10.1063/1.5001998} {\bibfield  {journal} {\bibinfo  {journal} {Biomicrofluidics}\ }\textbf {\bibinfo {volume} {11}},\ \bibinfo {pages} {054115} (\bibinfo {year} {2017})}\BibitemShut {NoStop}%
\bibitem [{\citenamefont {Settnes}\ and\ \citenamefont {Bruus}(2012)}]{settnes2012forces}%
  \BibitemOpen
  \bibfield  {author} {\bibinfo {author} {\bibfnamefont {M.}~\bibnamefont {Settnes}}\ and\ \bibinfo {author} {\bibfnamefont {H.}~\bibnamefont {Bruus}},\ }\bibfield  {title} {\bibinfo {title} {Forces acting on a small particle in an acoustical field in a viscous fluid},\ }\href {https://doi.org/10.1103/physreve.85.016327} {\bibfield  {journal} {\bibinfo  {journal} {Phys. Rev. E}\ }\textbf {\bibinfo {volume} {85}},\ \bibinfo {pages} {016327} (\bibinfo {year} {2012})}\BibitemShut {NoStop}%
\end{thebibliography}%
\end{document}